\documentclass{IEEEtran}
\usepackage{multirow}
\usepackage{cite}
\usepackage{amsmath,amssymb,amsfonts}
\usepackage{graphicx}
\usepackage{textcomp,nicefrac}
\usepackage{subfigure}
\usepackage{multicol}
\usepackage{subfig}
\usepackage{color}
\def\BibTeX{{\rm B\kern-.05em{\sc i\kern-.025em b}\kern-.08em
T\kern-.1667em\lower.7ex\hbox{E}\kern-.125emX}}
\begin{document}
\title{A novel measurement method for SiPM external crosstalk probability at low temperature }
\author{Guanda Li, Lei Wang, Xilei Sun, Fang Liu, Cong Guo, Kangkang Zhao, Lei Tian, Zeyuan Yu, Zhilong Hou, Chi Li, Yu Lei, Bin Wang, Rongbin Zhou
\thanks{
This work was supported by the National Key Research and Development Program (2022YFB3503600) and State Key Laboratory of Particle Detection and Electronics Fund (SKLPDE202310).}
\thanks{Lei Wang, is with School of University of Chinese Academy of Sciences, BeiJing, 100049, China (e-mail: leiwang@ihep.ac.cn).}
\thanks{Xilei Sun, Cong Guo, Lei Tian, Zeyuan Yu and Zhilong Hou are with Institute of High Energy Physics, Chinese Academy of Sciences, BeiJing, 100049, China (e-mail: sunxl@ihep.ac.cn; guocong@ihep.ac.cn; yuzy@ihep.ac.cn; houzl@ihep.ac.cn).}
\thanks{Guanda Li, Kangkang Zhao, Chi Li, Yu Lei, Bin Wang and Rongbin Zhou are with Institute of High Energy Physics, Chinese Academy of Sciences, BeiJing, 100049, China (e-mail: 
liguanda@ihep.ac.cn; zhaokk@ihep.ac.cn; lichi@ihep.ac.cn; leiyu@ihep.ac.cn; binwang@ihep.ac.cn; zhourb@ihep.ac.cn).}
\thanks{Fang Liu, is with School of Nuclear Science and Engineering, North China Electric Power University, Beijing 102206, China (e-mail: liuf@ncepu.edu.cn).}}
\maketitle

\begin{abstract}
Silicon photomultipliers (SiPMs) are being considered as potential replacements for conventional photomultiplier tubes (PMTs).
However, a significant disadvantage of SiPMs is crosstalk (CT), wherein photons propagate through other pixels, resulting in secondary avalanches. 
CT can be categorized into internal crosstalk and external crosstalk based on whether the secondary avalanche occurs within the same SiPM or a different one. 
Numerous methods exist for quantitatively estimating the percentage of internal crosstalk (iCT). 
However, external crosstalk (eCT) has not been extensively studied.

This article presents a novel measurement method for the probability of emitting an external crosstalk photon during a single pixel avalanche, using a setup involving two identical SiPMs facing each other, and without the need for complex optical designs. The entire apparatus is enclosed within a stainless steel chamber, functioning as a light-tight enclosure, and maintained at liquid nitrogen temperature. The experimental setup incorporates two Sensl J-60035 SiPM chips along with two 0.5-inch Hamamatsu Photonics (HPK) VUV4 S13370-6050CN SiPM arrays. 
The findings show a linear relationship between the probability of emitting an external crosstalk photon and the SiPM overvoltage for both SiPM samples.
Surprisingly, this novel measurement method also rovides measurements of the SiPM photon detection efficiency (PDE) for eCT photons at low temperature.
\end{abstract}


\section{Introduction}
\label{sec:introduction}
\IEEEPARstart{S}{ilicon} Photomultipliers (SiPMs) have emerged as a compelling solution for photodetection, capable of detecting single photons in various applications, such as particle physics, medical imaging, and distance measurement. Unlike the widely utilized Photomultiplier Tubes (PMTs), SiPMs operate at lower voltages, making them more suitable for low-temperature operation, where their performance is competitive. They are characterized by excellent single-photon resolution and low radioactivity \cite{1}. These photodetectors offer exceptional photon detection efficiency (PDE) not only in the visible and infrared wavelength ranges but also in the vacuum ultraviolet (VUV) wavelength range\cite{2}.

The single-photon detection ability of SiPMs is attributed to their exceptionally high gain, as a single electron-hole pair can trigger an avalanche resulting in charge amplification in the range of 10${^5}$ to 10${^7}$ \cite{3}. When a photon triggers an avalanche within a pixel,  a current pulse is driven to the external circuit with an amplitude proportional to the number of hit pixels. By computing differences in pulse height or the sum of pulse areas, photon counting and single-photon energy spectra can be obtained. However, an unfortunate byproduct of the avalanche generation process is the emission of secondary photons \cite{4}\cite{eCT1}. Those secondary photons can transfer between cells and create a secondary avalanche.

In general, secondary photons in SiPMs can lead to three main processes, according to the location of cells absorbing them \cite{CT1,CT2,Optical Crosstalk}. (1) If the secondary photon is absorbed by the neighboring Single Photon Avalanche Diode (SPAD) and triggers an additional avalanche within the same SiPM, it is referred to as internal optical crosstalk (iCT) \cite{iCT}. 
More specifically, iCT can further be divided into direct crosstalk and delayed crosstalk based on whether the electron is produced in the depleted region or in the undepleted region. 
More information can be found in \cite{CT_definition}.  
(2) Secondary photons can pass through the SiPM window, escape, and be detected by external photo-sensors, such as another SiPM facing the first one, known as external crosstalk (eCT) \cite{CT3}. 
(3)The 
final correlated effect is 
optically-induced afterpulsing (AP) \cite{AP}, resulting from electrons in the undepleted region within the same cell\cite{CT_definition}. 
As the result, optical-induced AP usually appears behind the primary pulse after hundreds of nanoseconds to microseconds because of the recharge time of SPADs.
As a result, optically-induced AP typically occurs after the primary pulse, ranging from hundreds of nanoseconds to microseconds, due to the recharge time of SPADs

These correlated effects could significantly impact experiments aiming to achieve a large photon-sensitive area using SiPM readout, as detailed in the Darkside and nEXO references \cite{DarkSide-20k}\cite{nEXO}. 
First of all, the light yield of the detector is probably overestimated since single incident photons would generate a signal with the integral area larger than 1 PE (photo-electron), because of the presence of iCT, eCT or AP. 
To estimate the corrected light yield, the knowledge of the probability of the three correlated effects is important. In addition to light yield overestimation, other negative effects need to be considered. 
For instance, the presence of AP can impede the n/$\gamma$ pulse shape discrimination in the background rejection analysis when using the prompt fraction method. 
Moreover, a high percentage of iCT and eCT would degrade the energy resolution. Additionally, eCT also generates unavoidable coincidental background in low threshold experiments, as eCT signals always occur simultaneously.

Overall, for any experiment employing large area SiPM arrays, the understanding of SiPM correlated noises is mandatory. The iCT probability is calculated by $P_{iCT}=\frac{N(1.5~PE)}{N(0.5~PE)}$ \cite{PDiCT}, where $N(1.5PE)$ is the dark count rate (DCR) with 1.5 PE threshold and $N(0.5PE)$ is the DCR with 0.5 PE threshold. 
Optical-induced AP could also be estimated using a similar method, given that the height of an afterpulse is the same as that of a single photoelectron. 
However, external crosstalk is not easily assessed. 
The typical method consists of collecting the photons emitted from the operating SiPM directly by employing sensitive optical equipment \cite{CT1}. 
For example, TRIUMF developed a novel device to estimate eCT, using a microscope, a spectrometer, and a CCD camera to measure the emitted light from SiPMs and the rate of secondary photons from each avalanche electron \cite{1}. They reported that the number of emitted photons per charge carrier in one avalanche is approximately $5\times10^{-6}$ photons per electron. 

This paper introduces a practical method for measuring the probability of eCT photon emission per single-pixel avalanche, utilizing two identical SiPM samples facing each other. The study tested two types of SiPM: SensL J-60035 and Hamamatsu Photonics (HPK) VUV4 S13370-6050CN. The entire experiment took place in a stainless steel chamber, serving as a dark box at liquid nitrogen temperature. Both SiPM types are considered excellent devices for low-temperature experiments\cite{PDiCT}\cite{LiuF}.

\section{Method}\label{sec:section2}
\begin{figure}[htbp]  

	\subfigure[]{
	\begin{minipage}[t]{0.8\linewidth}
	\centering     
    \includegraphics[width=1.0\linewidth]{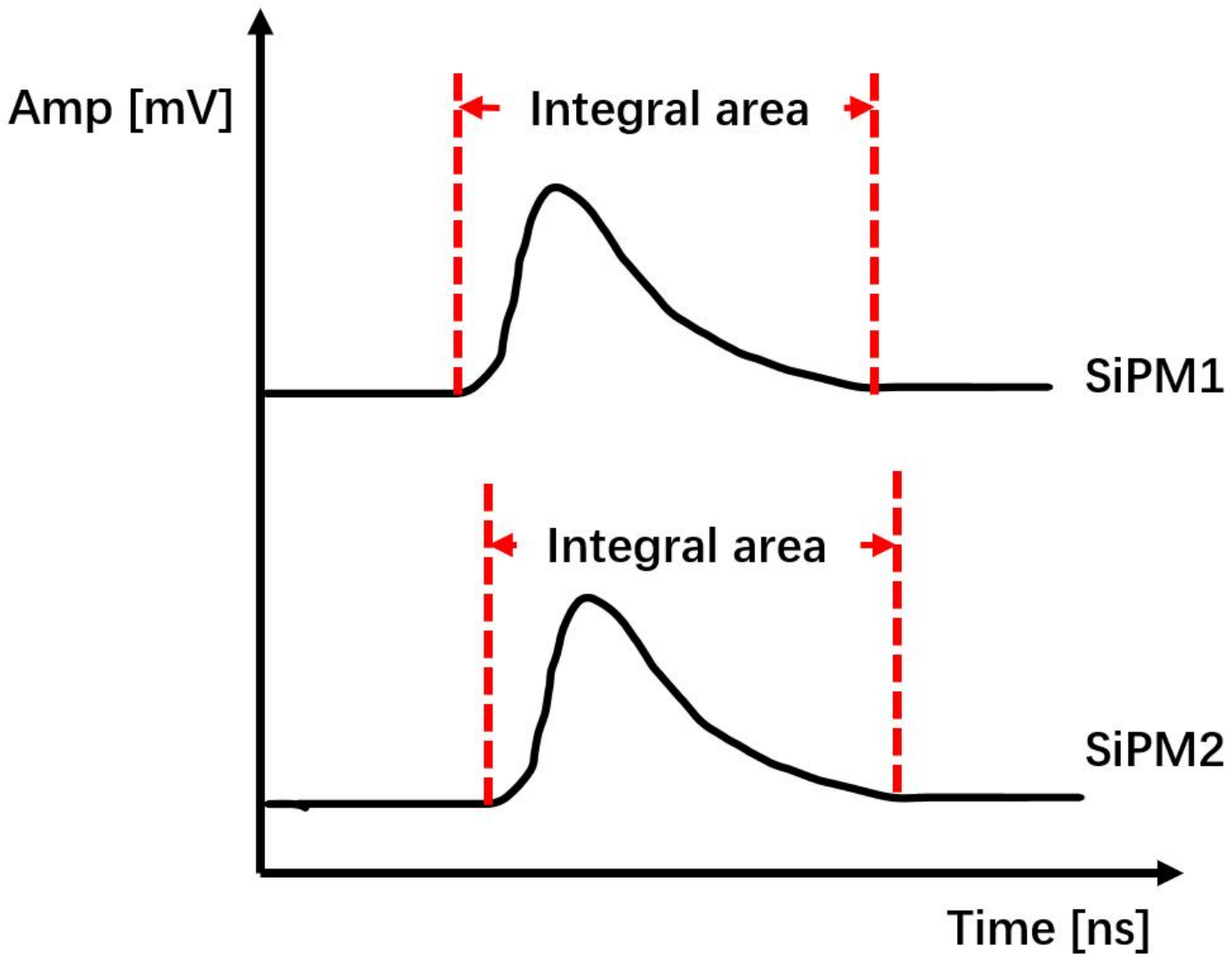}
	\end{minipage}
}

    \subfigure[]{
	\begin{minipage}[t]{0.8\linewidth}
	\centering     
	\includegraphics[width=1.0\linewidth]{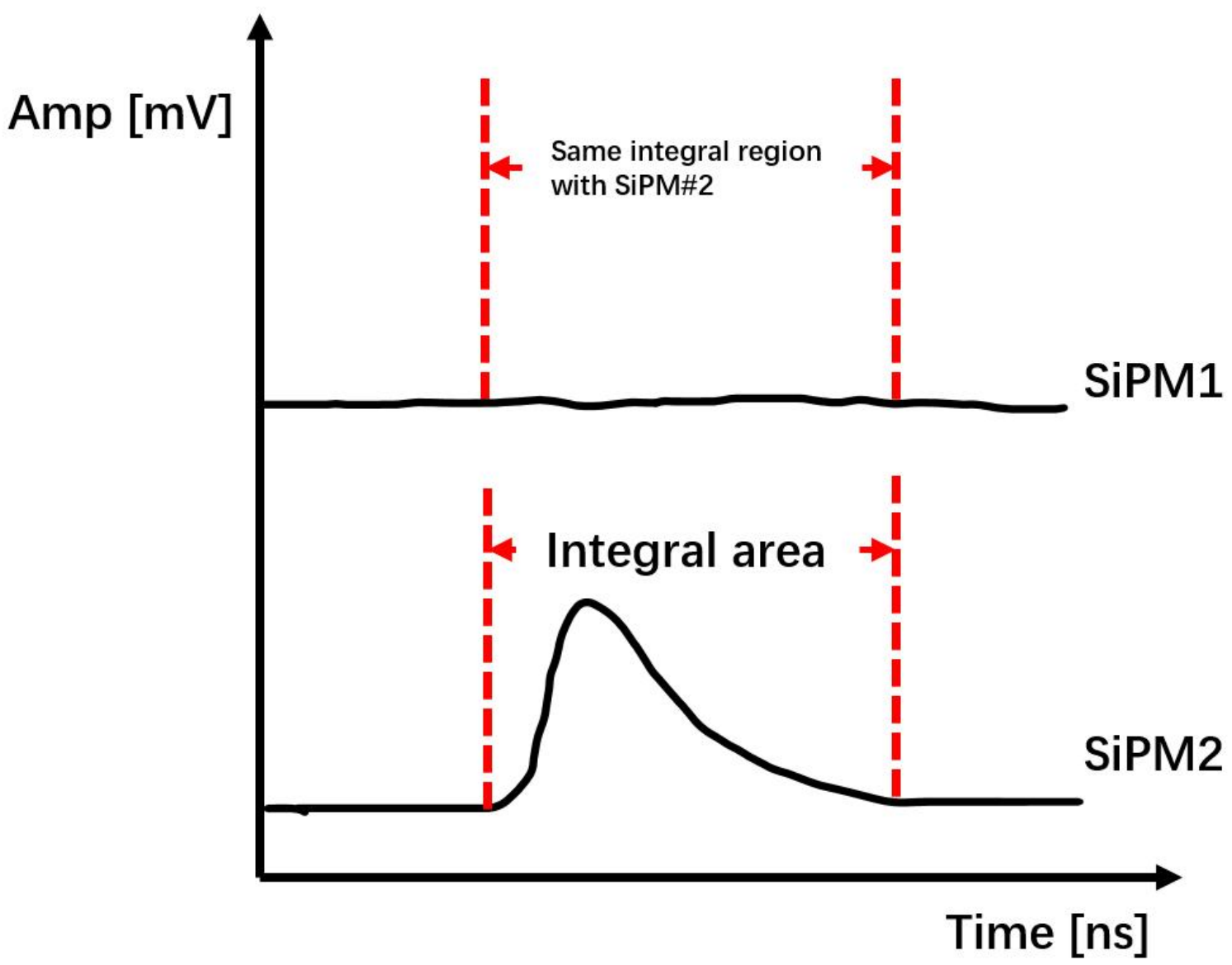}
	\end{minipage}
}
	\caption{Illustrations of the software selection for two SiPM samples (SiPM1 and SiPM2)}.
	\label{Illustration}           
\end{figure}

\begin{figure}[t]
\centerline{\includegraphics[width=3.5in]{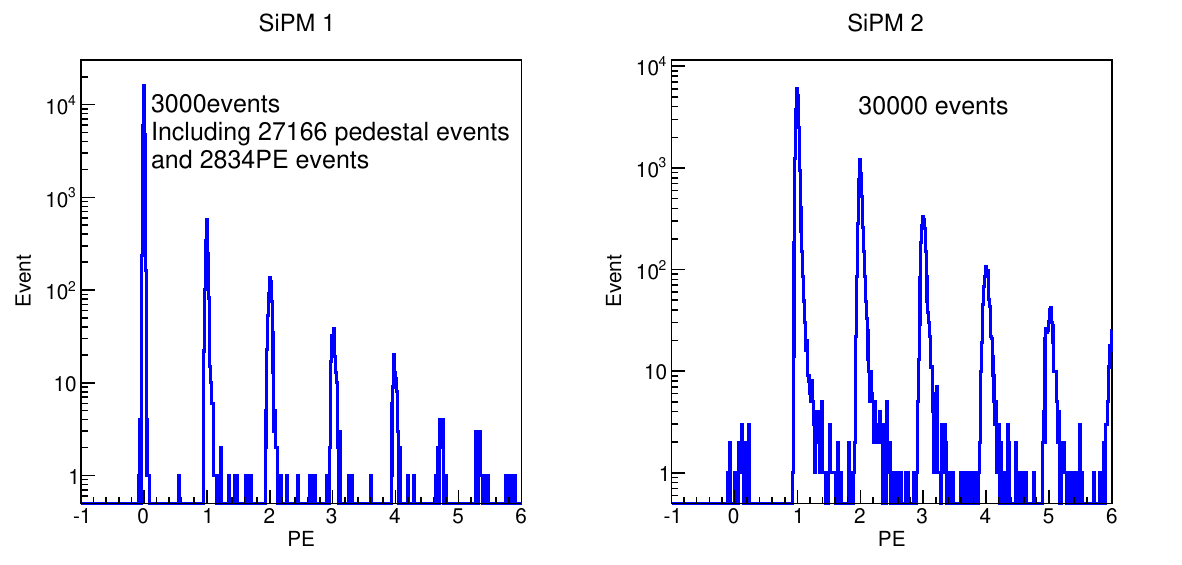}}
\caption{\label{PE distribution 1} {PE distributions of two Sensl J-60035 SiPMs facing each other. Both graphs present 30000 dark signals in the trigger mode where only SiPM2 was triggered with the threshold of 0.5~PE.}}
\end{figure}

When two identical SiPMs are positioned facing each other in a light-tight enclosure at a constant temperature and in close proximity, the eCT photons emitted from one SiPM are anticipated to interact with the surface of the opposing SiPM. The probability of secondary avalanches in the opposing sensor is subsequently determined by the PDE.
In the trigger mode, where only one SiPM is triggered with a 0.5 PE height threshold, while recording dark signals of both SiPMs simultaneously, two distinct dark signal distributions are observed.
For instance, let's consider two SiPMs labeled as SiPM1 and SiPM2, where SiPM2 serves as the emitter and SiPM1 as the detector. Only SiPM2 is triggered. 
The two SiPMs are facing each other with the same overvoltage. In this scenario, the photo-electron distribution of SiPM2 should be free of pedestal due to the deployment of the 0.5 PE height trigger. 
Subsequently, a software algorithm was used to search for coincidence signals in SiPM1 within a time gate T$_0$+dT, where dT=50ns and T$_0$ is the trigger time for SiPM2. If there is a PE signal in the time window, we integrate over the area of the PE signal (Fig.~\ref{Illustration}(a)) of SiPM1. On the contrary, if there is nothing but baseline in the gate, we integrate over the same region of SiPM2 in SiPM1(Fig.~\ref{Illustration}(b)). In this trigger mode, the PE distribution of SiPM1 has a majority of pedestal. For example, Fig~\ref{PE distribution 1} displays the PE distributions of two SensL SiPMs in a dark environment at 4.5V overvoltage and liquid nitrogen temperature.

\begin{figure}[t]
\centerline{\includegraphics[width=3.5in]{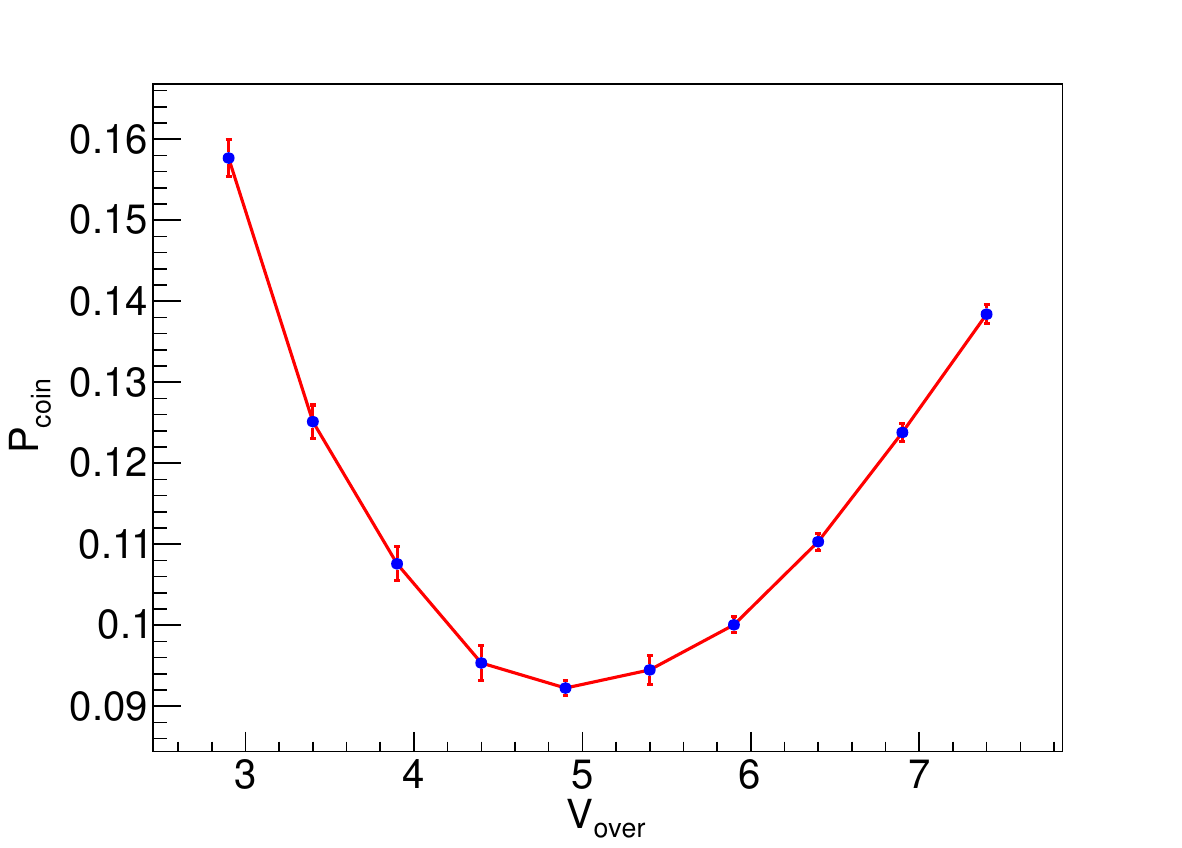}}
\caption{\label{Sensl Pcoin function}{  Coincidence probability $P_{coin}$ versus $V_{over}$ of SiPM2 measured by two Sensl J-60035 SiPMs facing each other, }with SiPM1 $V_{over}$ fixed at 4.4V.} 
\end{figure}

The coincidence probability $P_{coin}$ is defined as the ratio of PE events on SiPM1 in coincidence with PE events on SiPM2 to triggered PE events on SiPM2, resulting in $P_{coin}$ = 2834/30000 = 9.45$\%$, as shown in Fig~\ref{PE distribution 1}. 
Initially, $P_{coin}$ reflects the eCT level of both SiPM1 and SiPM2, as eCT photons emitted from the dark noise of SiPM2 are then received by SiPM1, and vice versa, both contributing to the coincidence probability. 
If this is accurate, an increase in the overvoltage of SiPM2 would be expected to lead to a higher $P_{coin}$, as the overvoltage enhances the SiPM gain, resulting in more charge carriers in avalanches.
Theoretically, a higher number of charge carriers in a single avalanche corresponds to a higher eCT probability.
In other words, $P_{coin}$ as a function of the SiPM overvoltage $V_{over}$ is expected to be a monotonically increasing function.
However, our measurement indicates that $P_{coin}$ does not increase consistently with $V_{over}$ for SiPM2. Fig~\ref{Sensl Pcoin function} shows $P_{coin}$ as a function of $V_{over}$ for SiPM2. 
It is worth noting that accidental coincidences of dark signals are disregarded, as the DCR of the SiPM is approximately 0.2 Hz/mm$^2$ at liquid nitrogen temperature. 
At this dark noise level, the accidental coincidence count rate is five to six orders of magnitude lower than the eCT count rate in the face-to-face setup.

 The unusual trend in the behavior of $P_{coin}$ as a function of $V_{over}$ for SiPM2 can be attributed to the variation in SiPM2 DCR.  
 The SiPM PDE and DCR both increase with increasing $V_{over}$ \cite{Hamamatsu}\cite{OnsemiDS}. 
 Taking these factors into consideration, it can be anticipated that $P_{coin}$ will be:

\begin{small}
\begin{equation}
\begin{aligned}
    P_{coin} = \frac{(P_{eCT_{2}}\times PDE_{1}\times DCR_{2})+(P_{eCT_{1}}\times PDE_{2}\times DCR_{1})}{DCR_{2}+(P_{eCT_{1}}\times PDE_{2}\times DCR_{1})} 
		\label{Pcoin function}
\end{aligned}
\end{equation}
\end{small}

As previously mentioned, $P_{coin}$ defines the probability of  coincidence at a specific trigger mode, where SiPMs face each other, but only SiPM2 is triggered. 
$P_{ect}$ represents the probability of a photon to be emitted from a SiPM surface under a specific temperature.
$PDE$ is the photon detection efficiency and $DCR$ is the dark count rate. 
The subscript number corresponds to the specific SiPM sample.
Importantly, even in the ideal case where two SiPMs are identical, the function remains asymmetrical because only SiPM2 is triggered. Therefore, regardless of the value of $V_{over}$, the denominator is always dominated by $DCR_{2}$.
When $V_{over}$  for SiPM2 is increasing, multiple components in Eq.~\ref{Pcoin function} change synchronously, like $P_{eCT_{2}}$, $DCR_{2}$ and $PDE_{2}$. 
The rapid increase of $DCR_{2}$ in the denominator of Eq.~\ref{Pcoin function} explains the abnormal trend of $P_{coin}$ in Fig.~\ref{Sensl Pcoin function}. 
When SiPM2 operates at a low voltage, $DCR_{2}$ is significantly smaller than $DCR_{1}$.
The cross-talk is primarily driven by SiPM1 to SiPM2, meaning that the numerator is mainly influenced by the second term, which remains relatively constant with Vover.
However, the denominator is affected by DCR2, which is rapidly increasing and growing at a faster rate than the numerator, resulting in a decrease in $P_{coin}$. 
When the dark count rate of SiPM2 increases and the crosstalk from SiPM2 to SiPM1 becomes dominant, the numerator is primarily influenced by the first term. Subsequently, the numerator begins to increase at a faster pace than the denominator, leading to an increase in in $P_{coin}$.
We have obtained the parametric equation of  $P_{coin}(V_{over_{2}})$ using $DCR$, $PDE$ and $P_{eCT}$. 
This implies that by fitting function in Fig.~\ref{Sensl Pcoin function} using Eq.~\ref{Pcoin function}, $P_{eCT}$ versus $V_{over}$ can be estimated, provided that we know the functions $DCR$($V_{over}$) and  $PDE$ ($V_{over}$). 
In fact, sometimes the precise $PDE$ function may not be necessary as it could be fitted together with $P_{eCT}$,
a topic that will be quantitatively explored in subsequent sections of this paper.

This section provides a concise introduction to the method.
$P_{coin}(V_{over_{2}})$ is measured and  Eq.~\ref{Pcoin function} is used to fit the data, indirectly estimating $P_{eCT}(V_{over})$. Detailed results and estimations are presented in section~\ref{sec:section4}.

\section{Experimental setup}\label{sec:section3}

\begin{figure}[t]
\centerline{\includegraphics[width=2.5in]{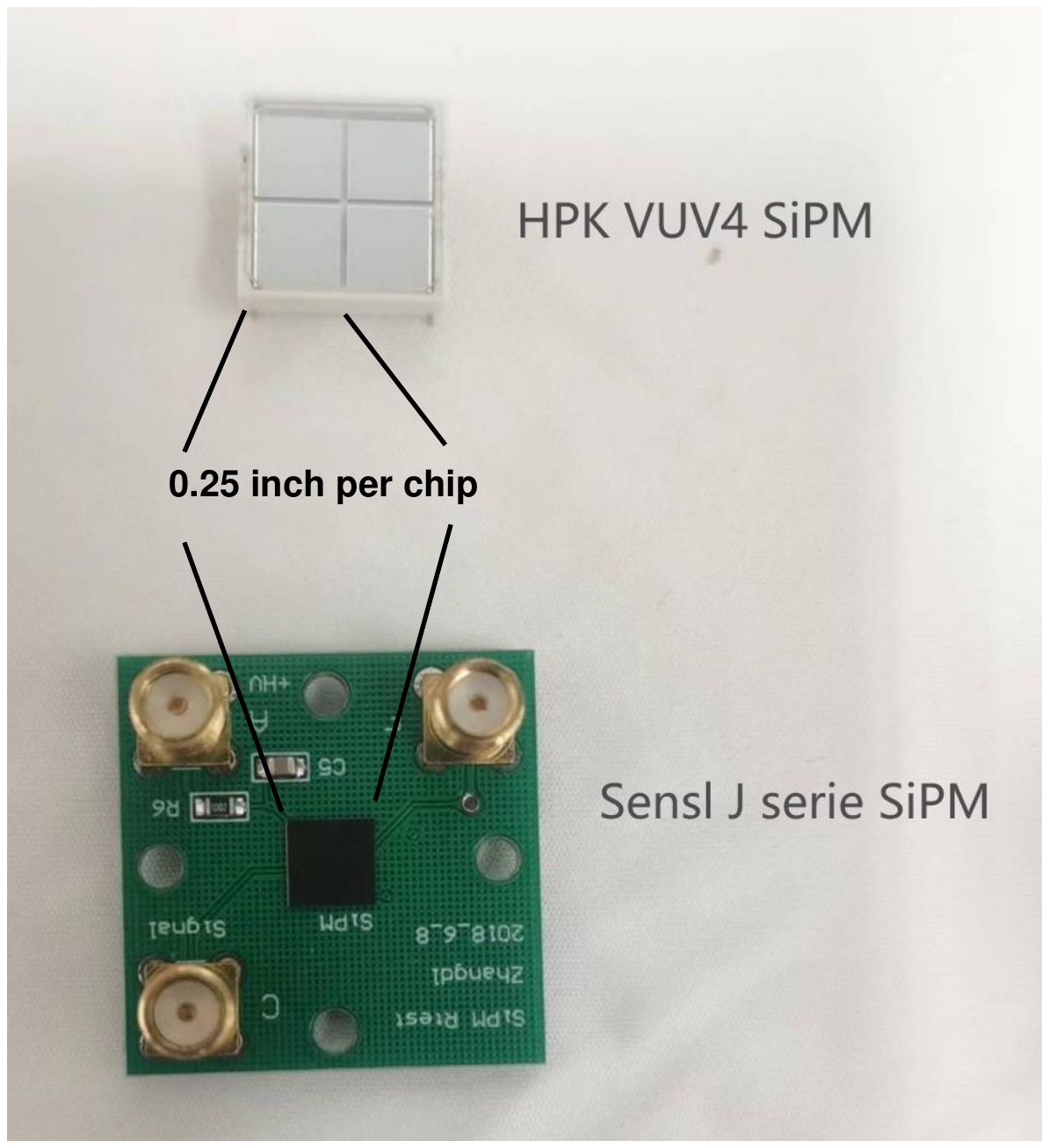}}
\caption{\label{SiPM_photo} Photographs of SiPM samples. HPK VUV4 S13370-6050CN SiPM consists of four chips and Sensl J-60035 SiPM is a single chip. Both chip lengths are 0.25-inch.}
\end{figure}

Two SiPM samples were employed for the eCT studies,including two SensL J-60035 SiPM chips and two 0.5-inch Hamamatsu Photonics (HPK) VUV4 S13370-6050CN SiPM arrays, as depicted in Fig.~\ref{SiPM_photo}.
Previous research has demonstrated the stability of these SiPMs when operating in a cryogenic environment \cite{liquid argon}. 

\subsection{Low temperature system}

The entire measurement was conducted at the temperature of liquid nitrogen in order to reduce the dark count rate (DCR) of the SiPM,  significantly decreasing the accidental coincidence “”rate between the two channels. 
Additionally, the study is part of the research and development for a low-temperature Coherent Elastic Neutrino-Nucleus Scattering (CE$\nu$NS) detector utilizing SiPM readout.
Two SiPM devices were chosen as potential candidates for the detector.

\begin{figure}[htb]
\centering
\centerline{\includegraphics[width=3.5in]{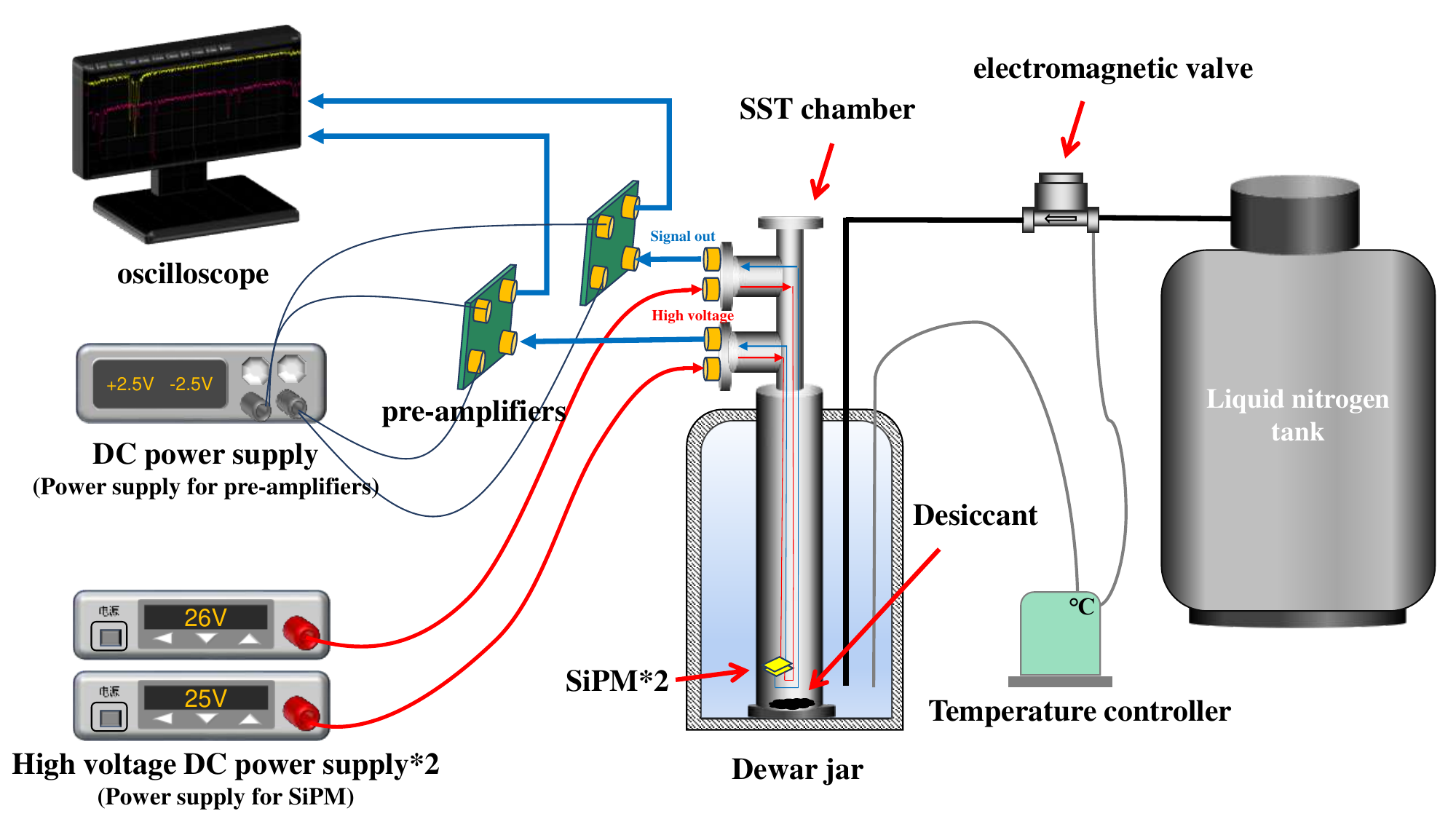}}
\caption{\label{SketchMap} Schematic diagram of the experiment.}
\end{figure}

The schematic diagram of the cryogenic system, shown in Fig~.\ref{SketchMap}, includes a liquid nitrogen tank, a dewar, and a stainless steel (SS) chamber. 
An electromagnetic valve was utilized to control the rate of liquid nitrogen inflow to the dewar.
The chamber does not need to be evacuated but must be kept dry. This can be achieved by purging the chamber with nitrogen. A more convenient method is simply using a couple of desiccant bags.

\begin{figure}[t]
\centerline{\includegraphics[width=2in]{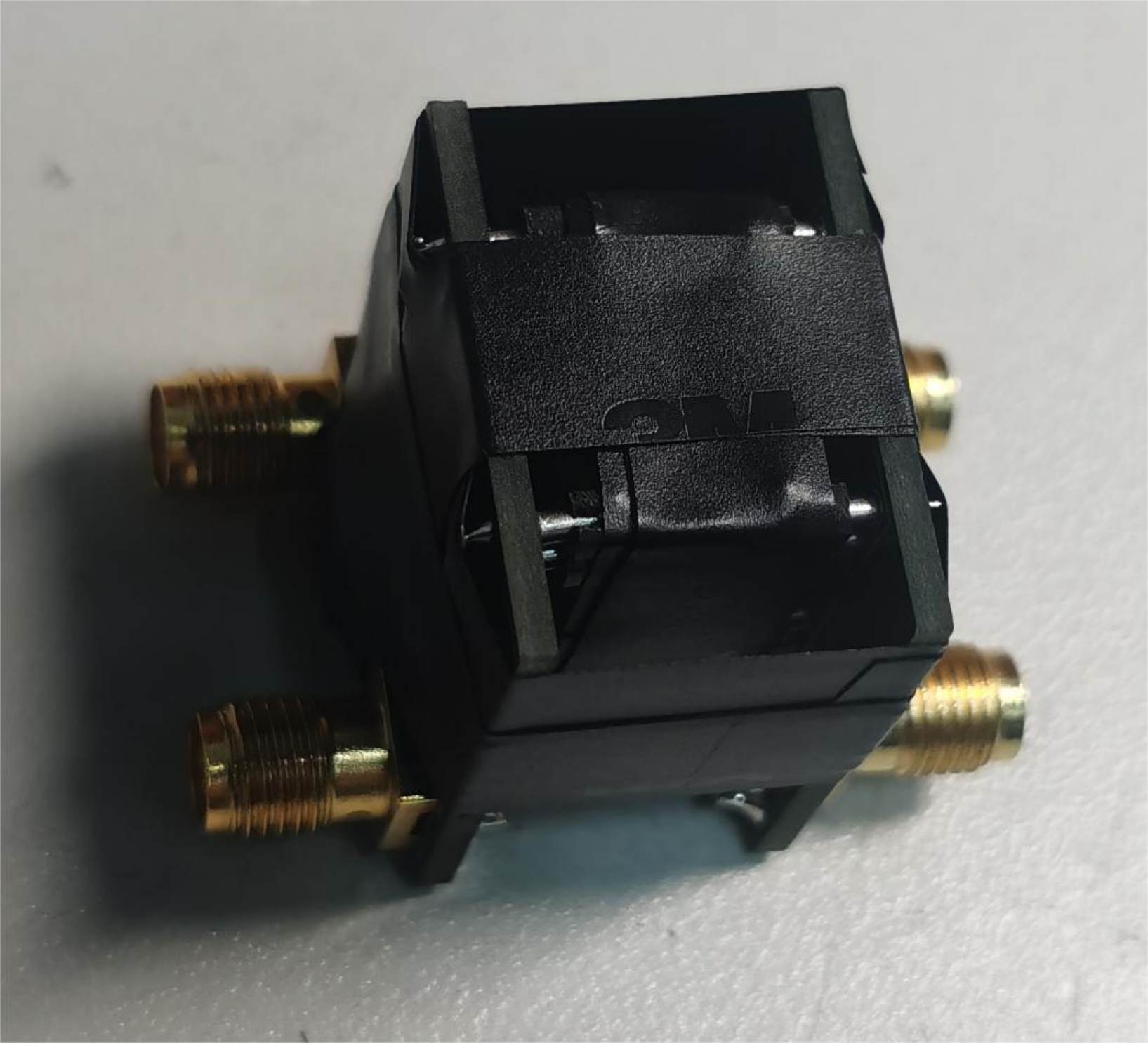}}
\caption{\label{VUV4} Photograph of two HPK VUV4 SiPMs facing each other and fixed with black tape}
\end{figure}

\subsection{SiPM and DAQ setups}
Two SiPMs of the same model are tightly attached together, facing each other. After wiring, the two SiPMs are fixed together with black tape, and then lightproofed by wrapping two layers of black cloth around them and fixing them with cable ties. To prevent condensation during the cooling process, the two SiPMs are placed into a sealed bag closed with cable ties. Fig.~\ref{VUV4} shows a photograph of two HPK VUV4 SiPMs facing each other and fixed in place with black tape. This setup ensures that the SiPMs are entirely unaffected by water vapor during the cooling process, thereby safeguarding the measurements.

The bias voltage and signal output pins of the SiPMs (shown in fig. 6) are connected to the feedthrough via 0.3-meter-long coaxial cables. The SiPM signals are driven to LMH6629 pre-amplifiers\cite{6629} for amplification before being sampled by an oscilloscope for data acquisition.

Two SiPM arrays were placed together in the SS chamber for testing. The operating voltage was provided by two DH 1765-4 DC power supplies. A RIGOL DP831A DC power supply was used to bias the two LMH6629 pre-amplifiers. A LeCroy 104Xs-A oscilloscope was used for data acquisition. The oscilloscope is in self-trigger mode, with the threshold set to 0.5PE amplitude. The time window is set to 5000 ns, and the data sampling rate of the oscilloscope is set to 500 MHz.

\section{eCT fitting}\label{sec:section4}

\subsection{Dependence on overvoltage} 

The principle of eCT estimation has been introduced in Sec.~\ref{sec:section2}. Some premises need to be observed for achieving the final fitting process. 
First of all, some terms in Eq.~\ref{Pcoin function} like $P_{eCT}$, $PDE$ and $DCR$ depend on  overvoltage, and those dependences should be obtained.

The SiPM DCR as a function of $V_{over}$ is defined as an exponential function with three coefficients (Eq.~\ref{DCR function}):
\begin{equation}
	DCR(V_{over}) = a_1e^{(a_2\times V_{over})}+a_3
	\label{DCR function}
\end{equation}

\begin{figure}[htbp]  

	\subfigure[DCR versus $V_{over}$ for the SensL SiPM sample.]{
	\begin{minipage}[t]{0.8\linewidth}
	\centering     
    \includegraphics[width=1.3\linewidth]{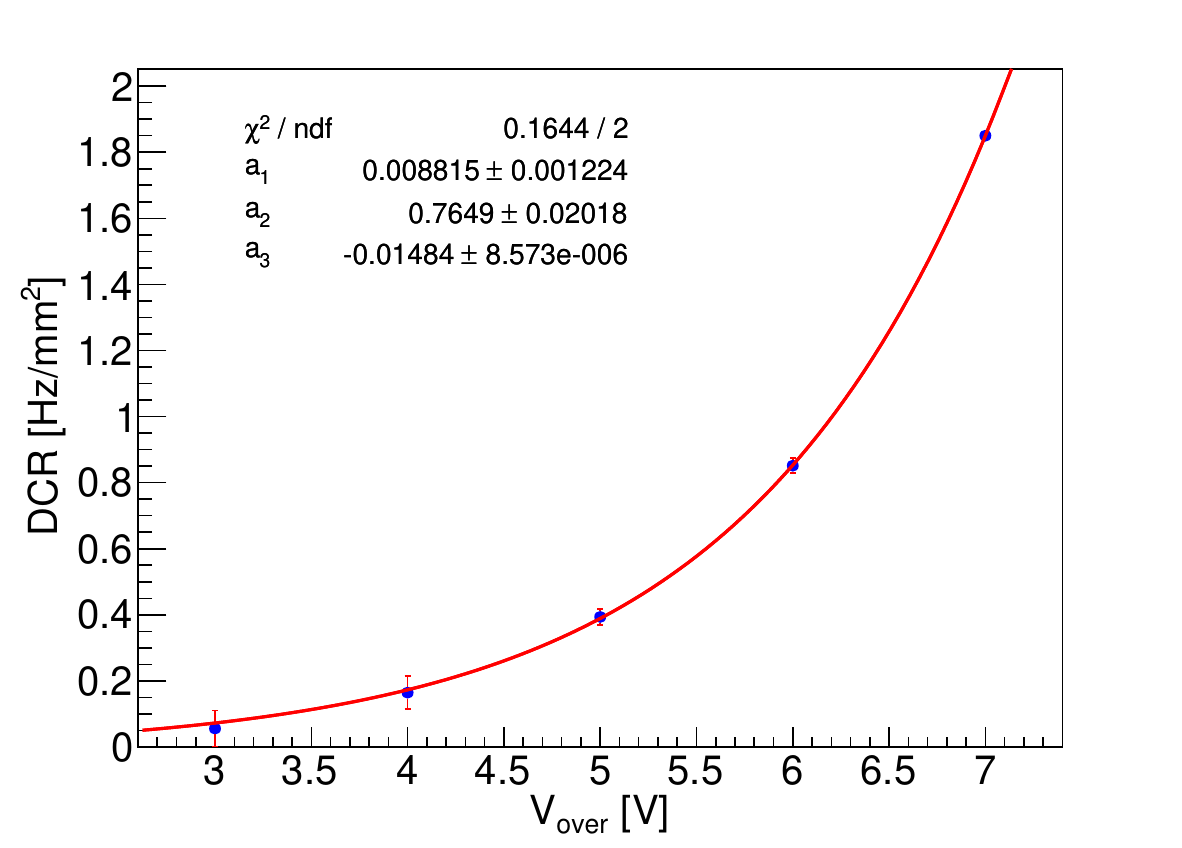}
	\end{minipage}
}

    \subfigure[DCR versus $V_{over}$ for the HPK VUV4 sample.]{
	\begin{minipage}[t]{0.8\linewidth}
	\centering     
	\includegraphics[width=1.3\linewidth]{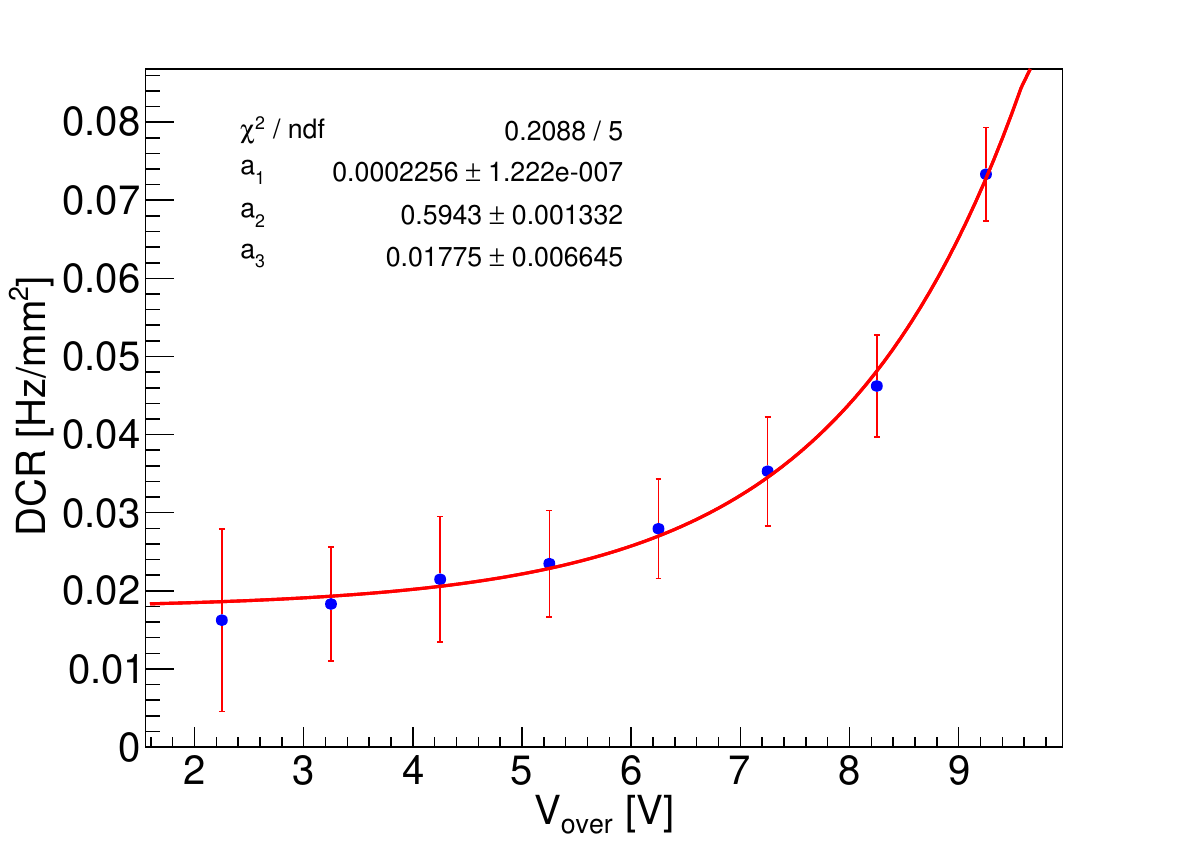}
	\end{minipage}
}
	\caption{ DCR versus $V_{over}$ for SensL SiPM (a) and for HPK SiPM (b).} Data are fitted as exponential functions with three parameters.  
	\label{dcr}           
\end{figure}

The coefficients in Eq. 2 are calculated by fitting the DCR data measured at dark environment under temperature of 76K~\cite{PDiCT}. Fig.~\ref{dcr} shows the fitting results of the DCR function for both SiPMs and the value of coefficients $a_1$, $a_2$ and $a_3$.

$P_{eCT}$ should be a linear function of $V_{over}$ :
\begin{equation}
	P_{eCT}(V_{over}) = k \times V_{over}
	\label{eCT function}
\end{equation}

The coefficient $k$ represents the variation of $P_{eCT}$ with respect to $V_{over}$, with units of V$^{-1}$. 
This functional form is reasonable because $P_{eCT}$ should be proportional to the SiPM gain, and the gain is proportional to $V_{over}$. As a consequence, in the operating region of the SiPM, $P_{eCT}$ is expected to increase linearly with $V_{over}$. Once $k$ is estimated, the probability of emitting one eCT photon per avalanche ($P_{eCT}$) at a certain voltage can be obtained.

$PDE$ as a function of $V_{over}$ is the most difficult dependence to be parameterized for several reasons. 
Firstly, there is no physical formula describing the dependence between $PDE$ and $V_{over}$, though they are correlated. Secondly, SiPM datasheet provided from manufacturers usually shows $PDE(V_{over})$ only at 420~nm and room temperature. It is expected that the eCT spectrum of SiPMs is a continuous spectrum from visible to infrared wavelength range \cite{eCT1}, which means we currently do not know the accurate PDE function for eCT photons because it may depend on temperature. 
By referring the article in \cite{PDE1},  $PDE(V_{over})$ is simply defined in Eq.~\ref{PDE_420nm function} with two parameters.  

\begin{equation}
	PDE(V_{over}) = PDE_{max}(1-e^{-V_{over}/(V_{bd}\times\alpha)})
	\label{PDE_420nm function}
\end{equation}

$V_{bd}$ is the breakdown voltage of the SiPM at a certain temperature. It is known in the analysis. $V_{bd}$ for HPK VUV4 SiPM and SensL SiPM is 42.3~V and 20.6~V at 76~K, respctively, according to our measurement. Funtion $PDE(V_{over})$ around 420~nm could be easily found in the literature and datasheet from manufacturers \cite{Hamamatsu}\cite{OnsemiDS}, but there is no reference about $PDE(V_{over})$ 
trend under conditions of the eCT spectrum. Thus, $PDE_{max}$ and $\alpha$ cannot be simply estimated like the  parameters in $DCR(V_{over})$.

We have obtained mathematical formulas for  $P_{eCT}(V_{over})$, $PDE(V_{over})$ and $DCR(V_{over})$ in Eq.~\ref{Pcoin function}. 
But it should be noticed that only the coefficients in $DCR(V_{over})$ are measured. Substituting Eq.~\ref{eCT function} and Eq.~\ref{PDE_420nm function} back into Eq.~\ref{Pcoin function}, we would find $P_{coin}$ with only two parameters to fit, which are $k \times PDE_{max}$ and $\alpha$. 
There is only one degree of freedom for the two parameters $k$ and $PDE_{max}$ in the fitting because they are always multiplied together, i.e., $P_{eCT}$ is always multiplied by $PDE$ in Eq.~\ref{Pcoin function}.  Considering the target of the paper to estimate the value of $k$, separating of two parameters $k$ and $PDE_{max}$ is necessary. 
The method could be found in Sec.~\ref{separating k and PDE} after the fitting is finished. 

\subsection{Correction of complex branching processes} 

The analysis, so far, has not considered complex optical processes between two SiPMs. For instance, in Fig.~\ref{PE distribution 1}, double PE events in SiPM2 should double the probability of emitting one eCT photon. Furthermore, eCT photons could transfer to the back surface of two SiPMs causing dependent large PE pulses. 
Thus, Eq.\ref{Pcoin function} is not a prefect equation to describe the dependence of $P_{coin}$ on $P_{eCT}$, PDE and DCR mathematically. The calculation of the coincidence probability needs to be corrected by eliminating the effect of iCT and those complicated branching optical effects. 

\begin{figure}[t]
\centerline{\includegraphics[width=3.5in]{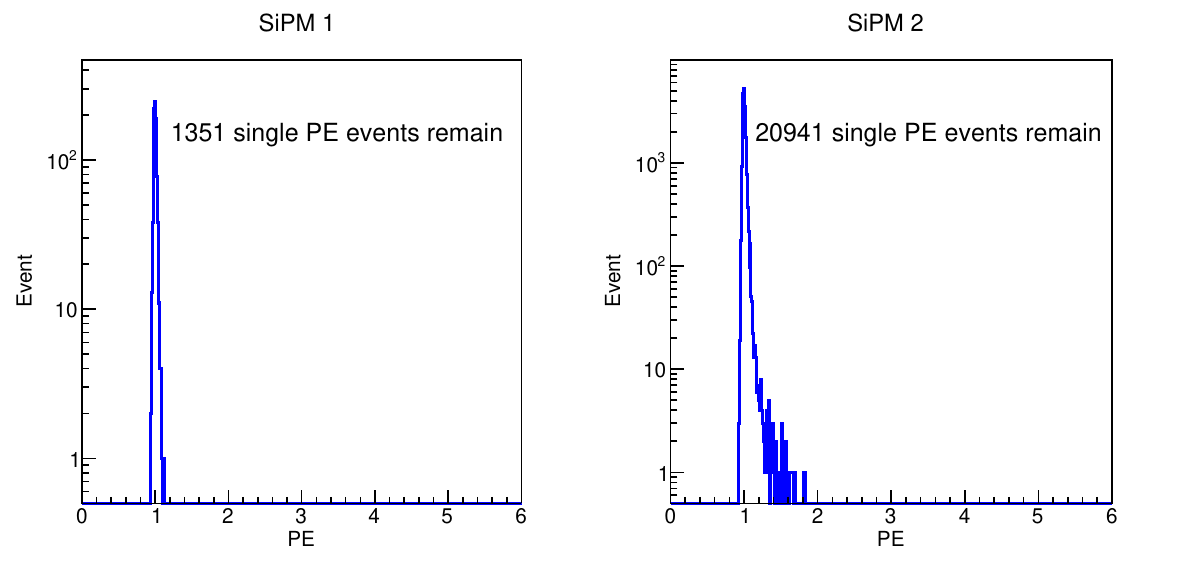}}
\caption{\label{PE distribution 2} A simplified model is used to eliminate the effect of iCT and dependent optical processes by cutting off all multi-PE events in Fig.~\ref{PE distribution 1}. Only the situation of one avalanche occuring and one eCT photon escaping the SiPM is accounted for.}
\end{figure}

A software selection procedure was developed to cut all multi-PE events in Fig.~\ref{PE distribution 1}.
Only single-PE events remain, as shown in fig.~\ref{PE distribution 2}, and then the corrected coincidence probability $P_{coin}$ was recalculated, resulting $P_{coin}$ = 1351/20941 = 6.45\% at $V_{over}$=4.5V for both SensL SiPMs.

In the direct cut condition, all complex dependent branching processes between the two SiPMs are eliminated, as these signals always manifest with multiple PE pulses. Only two optical processes are retained: "one dark noise photon in SiPM2 creates one avalanche, then emits one photon, which is received by SiPM1, causing another avalanche," and the reverse process. 
Eq.~\ref{Pcoin function} does not perfectly describe the simplified situation in Fig~\ref{Fitting Result} at present. It needs to be modified.
For example, an important point is that the selection lost two eCT processes: "one dark noise photon in SiPM2 creates one avalanche, then emits one photon, which is received by SiPM1, causing multiple avalanches because of the presence of iCT," and the reverse process. 
As the result, terms of $P_{eCT_2}$ in Eq.~\ref{Pcoin function} need to be multiplied by a $1-P_{iCT_1}$ term and terms of $P_{eCT_1}$ in Eq.~\ref{Pcoin function} need to be multiplied by a $1-P_{iCT_2}$ term. All the terms that require modification in Eq.~\ref{Pcoin function} are the following: 

\begin{eqnarray}\label{Eq_modifications}
\begin{split}
	P_{eCT_1}\rightarrow P_{eCT_1} \times (1-P_{iCT_2}) \\
	P_{eCT_2}\rightarrow P_{eCT_2} \times (1-P_{iCT_1}) 
\end{split}	
\end{eqnarray}

\begin{figure*}[htpb]
\begin{small}
\begin{equation}\label{Pcoin function modifed}
\begin{split}
	P_{coin} = \frac{(P_{ect_{2}}\times(1-P_{iCT_1})\times PDE_{1}\times DCR_{2})+(P_{ect_{1}}\times(1-P_{iCT_2})\times PDE_{2}\times DCR_{1})}{DCR_{2}+(P_{ect_{1}}\times(1-P_{iCT_2})\times PDE_{2}\times DCR_{1})}
\end{split}
\end{equation}
\end{small}
\end{figure*}

Substituting modified terms in Eq.~\ref{Eq_modifications} back into Eq.~\ref{Pcoin function} after applying the cut condition, a more  accurate expression of  $P_{coin}$ versus $V_{over}$ is obtained, as shown in Eq.~\ref{Pcoin function modifed}. 

\begin{figure}[htbp]  

	\subfigure[1-$P_{iCT}$ versus $V_{over}$ for the SensL SiPM sample.]{
	\begin{minipage}[t]{0.8\linewidth}
	\centering     
    \includegraphics[width=1.3\linewidth]{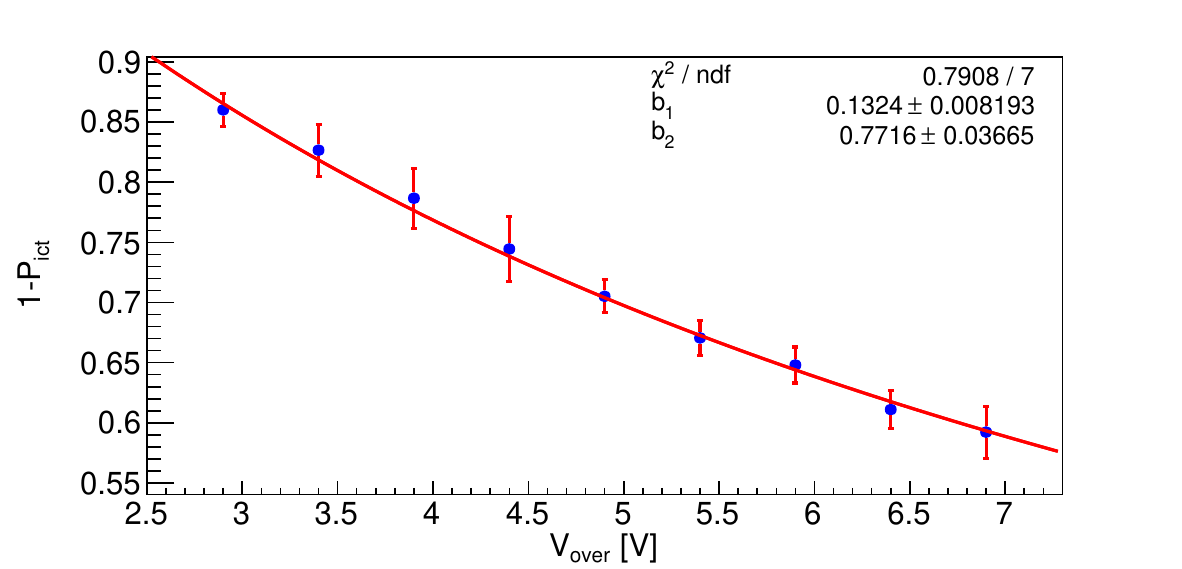}
	\end{minipage}
}

    \subfigure[1-$P_{iCT}$ versus $V_{over}$ for the HPK VUV4 sample.]{
	\begin{minipage}[t]{0.8\linewidth}
	\centering     
	\includegraphics[width=1.3\linewidth]{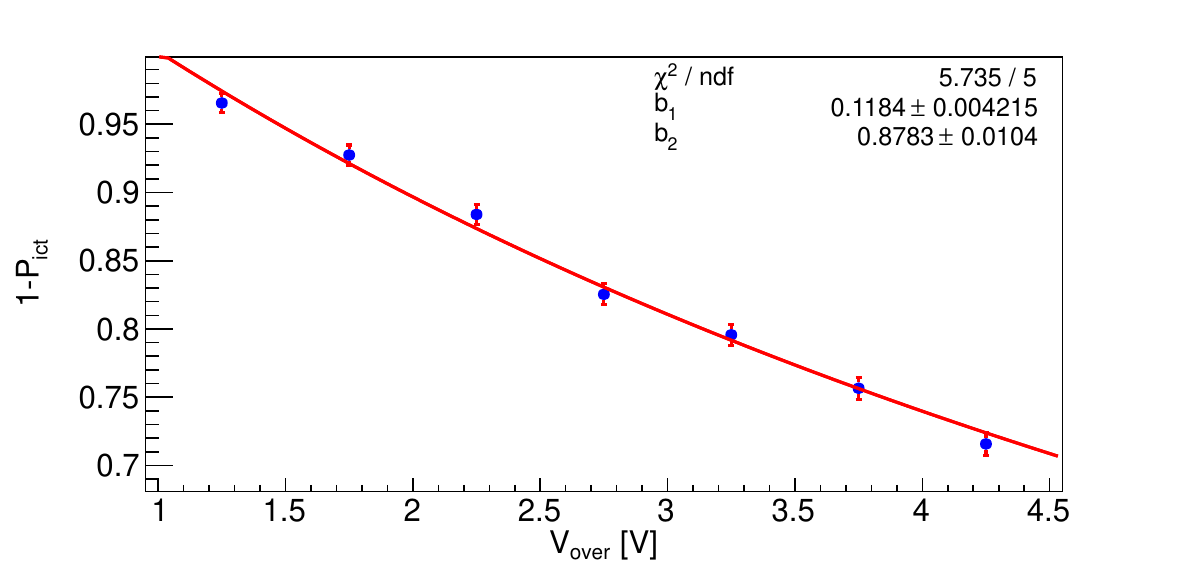}
	\end{minipage}
}
	\caption{1-$P_{iCT}$ as a linear function of $V_{over}$ measured under 76K dark environment for SensL and HPK VUV4 SiPM samples.}    
	\label{iCT}           
\end{figure}

$P_{iCT}=\frac{N(1.5~PE)}{N(0.5~PE)}$ was estimated by recording the dark signals of the SiPM in self-trigger mode without activating the other SiPM, then calculating the ratio of two and more PE pulses keeping temperature and geometry. $P_{iCT}$ as a function of $V_{over}$ for two SiPM samples was measured. The results are shown in Fig.~\ref{iCT}. In the analysis, $P_{iCT}$ versus $V_{over}$ is defined as an inverse proportional function with two parameters in Eq.~\ref{iCT function}. 

\begin{equation}
	1-P_{iCT}(V_{over}) = \frac{1}{b_1 \times V_{over}+b_2}
	\label{iCT function}
\end{equation}

\subsection{Fitting results} 
\begin{figure}[htbp]  

	\subfigure[$P_{coin}$ versus $V_{over}$ for SensL SiPM samples.]{
	\begin{minipage}[t]{0.8\linewidth}
	\centering     
    \includegraphics[width=1.3\linewidth]{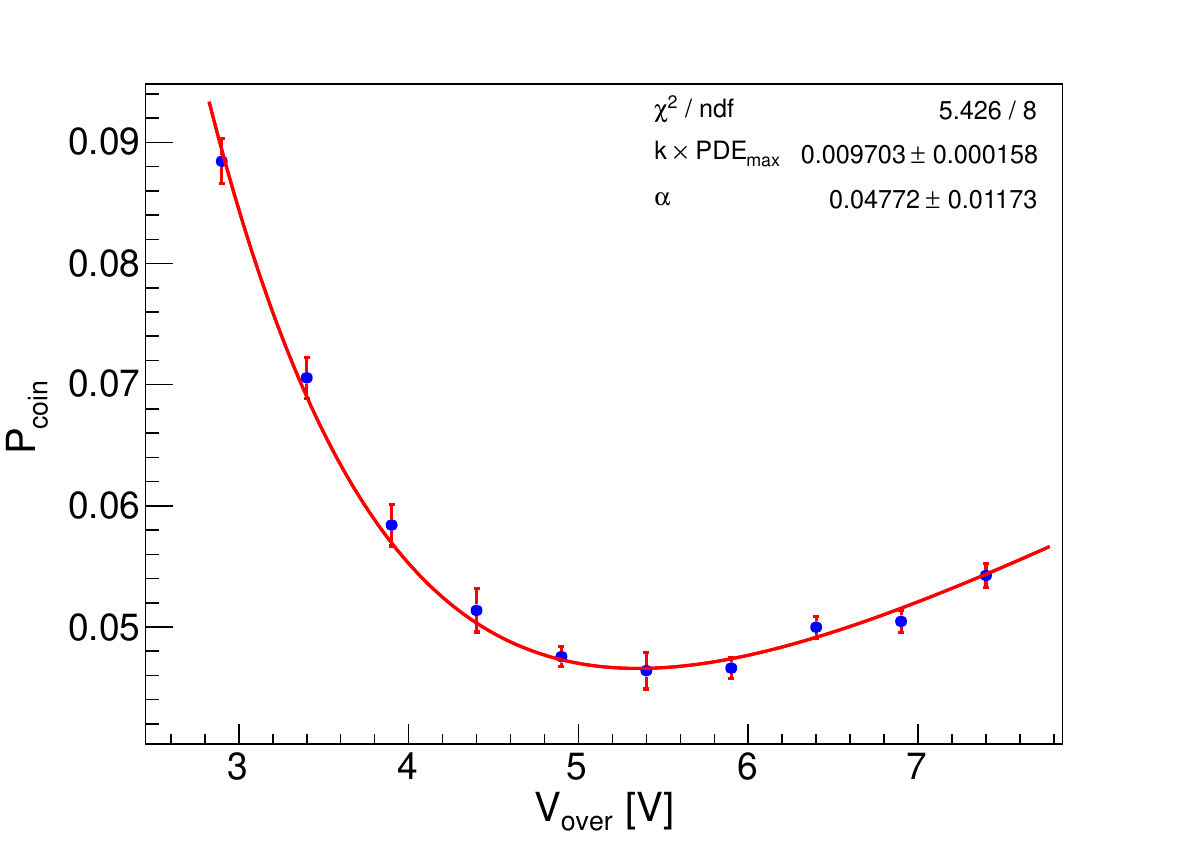}
	\end{minipage}
}

    \subfigure[$P_{coin}$ versus $V_{over}$ for HPK VUV4 samples.]{
	\begin{minipage}[t]{0.8\linewidth}
	\centering     
	\includegraphics[width=1.3\linewidth]{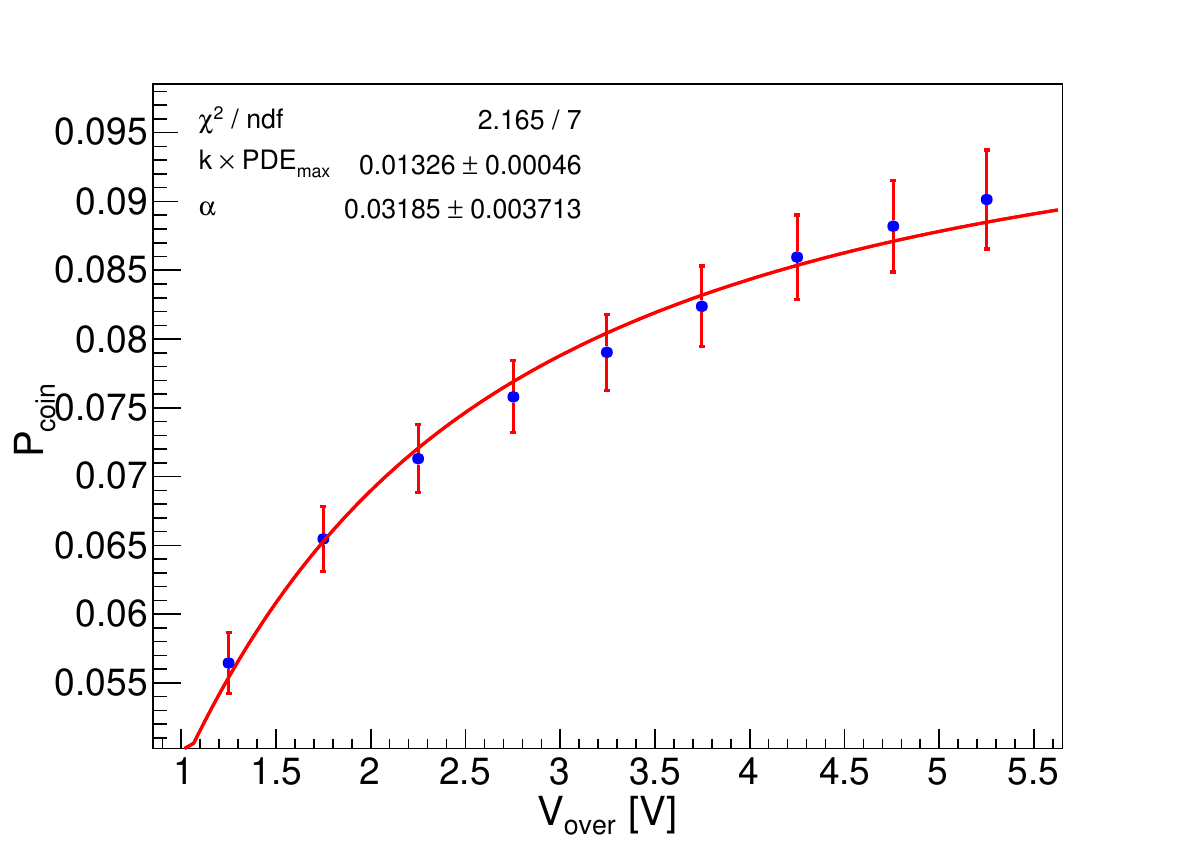}
	\end{minipage}
}
	\caption{ Coincidence probability $P_{coin}$ versus $V_{over}$ for SensL SiPM (a) and HPK SiPM (b). Two parameters are obtained from fitting: coefficients $k \times PDE_{max}$ and $\alpha$.}   
	\label{Fitting Result}           
\end{figure}

Based on Eq.~\ref{Pcoin function modifed}, the fitting equation for the coincidence probability is finally obtained, containing five identified parameters and two unknown parameters. Table~\ref{table.1} shows all identified parameters needed for the fitting. Fig.~\ref{Fitting Result} shows the fitting results of the coincidence probability $P_{coin}(V_{over})$ with two parameters, $k \times PDE_{max}$ and $\alpha$. Parameter $k$ comes from Eq.~\ref{eCT function}, which represents the only coefficient of $P_{eCT}$. 
Parameter $\alpha$ is defined in Eq.~\ref{PDE_420nm function} corresponding to the increase rate of PDE versus $V_{over}$.\cite{alpha} 

An interesting observation in Fig.~\ref{Fitting Result} is that the $P_{coin}$ function of the two SiPM samples shows a very different trend. 
However, both of them could be fitted by the same formula that only has two adjustable parameters.
The different behavior between the two types of SiPM samples is mainly due to the rise speed of function $DCR(V_{over})$.
In Fig.~\ref{dcr}(a), it is shown that the DCR of SensL SiPM increases much faster than that of HPK VUV4 SiPM, which causes the decrease of $P_{coin}$ in Fig.~\ref{Fitting Result}.

\begin{table*}[htb]
\caption{\\ Five fixed parameters are used in the fitting, including three parameters for the DCR function and two parameters for the $P_{iCT}$ function.
}
\centering
\renewcommand{\arraystretch}{1.5}
\normalsize
\begin{tabular}{cccc}
\hline
\multirow{2}{*}{SiPM type} & \multicolumn{2}{c}{$DCR(V_{over})$ parameters (Eq.~\ref{DCR function} and Fig.~\ref{dcr})} \\
 & $a_1$ & $a_2$ & $a_3$ \\
\hline
 HPK SiPM & 2.26$\times$10$^{-4}$  & 5.94$\times$10$^{-1}$ & 1.78$\times$10$^{-2}$ \\
 Sensl SiPM & 8.82$\times$10$^{-3}$ & 7.65$\times$10$^{-1}$ & -1.48$\times$10$^{-2}$\\
\hline
& \multicolumn{2}{c}{$P_{iCT}(V_{over})$ parameters (Eq.~\ref{PDE_420nm function} and Fig.~\ref{iCT function})} \\
 & $b_1$ & $b_2$ &  \\
\hline
 HPK SiPM & 1.18$\times$10$^{-1}$  & 8.78$\times$10$^{-1}$ &  \\
 Sensl SiPM & 1.32$\times$10$^{-1}$ & 7.72$\times$10$^{-1}$ &  \\
 \hline
\end{tabular}

\label{table.1}
\end{table*}

\subsection{disentangling $k$ from $PDE_{max}$}\label{separating k and PDE}

For now, we only calculated the product of $k$ and $PDE_{max}$ for two SiPM samples. $k$ and $PDE_{max}$ share one degree of freedom in the fitting.
If $PDE_{max}$ could be determined,
the parameter $k$, which represents the probability of emitting one photon per avalanche under a certain $V_{over}$, could be constrained. 
Since we already have the value of $\alpha$, another interesting byproduct is the $PDE$  function for eCT photons. 
Usually, it is difficult to estimate due to eCT photons being a continuous spectrum from the visible to the infrared wavelength range.

\begin{figure}[t]
\centerline{\includegraphics[width=3.5in]{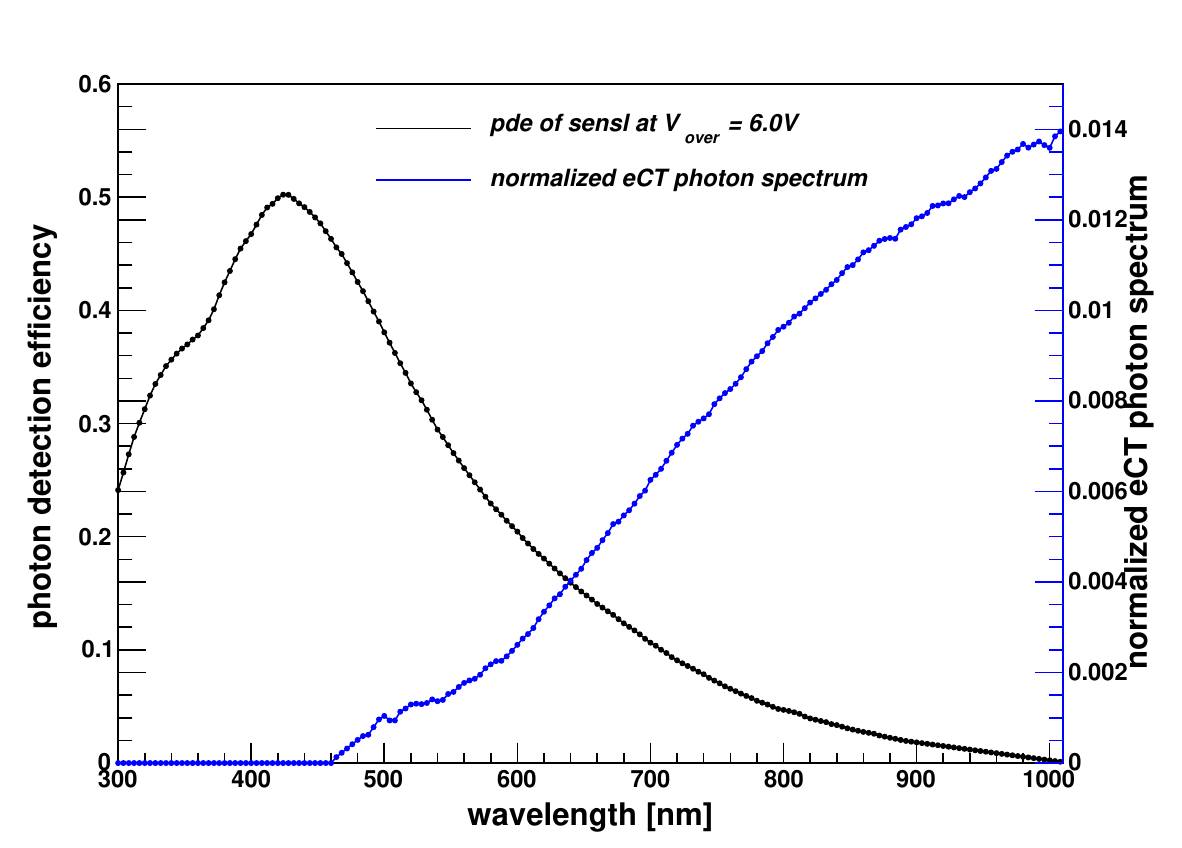}}
\caption{\label{Sensl_pde_pect} The black curve represents the PDE versus wavelength for the SensL SiPM sample at 420~nm and $V_{over}$ 6V, obtained from the SensL J-series datasheet \cite{OnsemiDS}. 
The blue curve represents the normalized eCT spectrum data for the VUV4 SiPM sample, extracted from \cite{1}.}
\end{figure}

Fig.~\ref{Sensl_pde_pect} displays the normalized eCT photon spectrum versus wavelength of the VUV4 SiPM and the PDE function versus wavelength of the SensL SiPM at 420 nm with $V_{over}$=6V. 
The normalized eCT photon spectrum data was obtained from \cite{1}, and the PDE data is obtained from the manufacturer datasheet \cite{OnsemiDS}. 
We also possess the accurate PDE function versus wavelength for the VUV4 SiPM at $V_{over}$=4V from the company, but we were asked not to disclose it. 
Mathematically, the PDE for the eCT photon spectrum could be calculated by integrating the normalized eCT photon spectrum and the PDE function versus wavelength.
However, there is no literature describing the eCT photon spectrum for SensL SiPM. Therefore, an assumption is made in the paper: the normalized eCT photon spectra are almost the same for different types of SiPM at a wavelength of 900 nm, even though their non-normalized spectra are expected to be different.
As a result, the PDE for SensL SiPM at $V_{over}$=6V and for HPK VUV4 SiPM at $V_{over}$=4V were calculated based on the normalized eCT photon spectrum of the HPK VUV4 SiPM. The calculated PDE values are 5.50$\pm$0.98\% for the SensL SiPM at  $V_{over}$=6V and 10.5$\pm$0.53\% for the HPK VUV4 SiPM at $V_{over}$=4V. 
The uncertainties in the PDE calculation come mainly from the error in PDE versus wavelength provided by the datasheets and the measured normalized eCT spectrum. 
Evaluating data errors from the  datasheet is challenging due to the absence of error bars in the graph. Additionally, the error associated with the normalized eCT spectrum is also minimal.
Finally, we assume an uncertainty of 5\% for the HPK SiPM PDE at $V_{over}$=4V. In contrast, the uncertainty for the SensL SiPM PDE at $V_{over}$=6V was set to 17.8\% for a different reason. 
Specifically, the assumption in the PDE calculation is that the normalized eCT photon spectra are the same for HPK VUV4 and SensL SiPMs.
However, this assumption is not grounded in fact.
In our previous work~\cite{1}, we estimated the difference in normalized eCT spectra between HPK VUV4 samples and FBK VUV-HD3 samples. 
They exhibited a difference of 17.8\%. 
Consequently, we set the uncertainty of the PDE calculation for the SensL SiPM at $V_{over}$=6V to 17.8\%.

\begin{figure}[htbp]
\centerline{\includegraphics[width=3.5in]{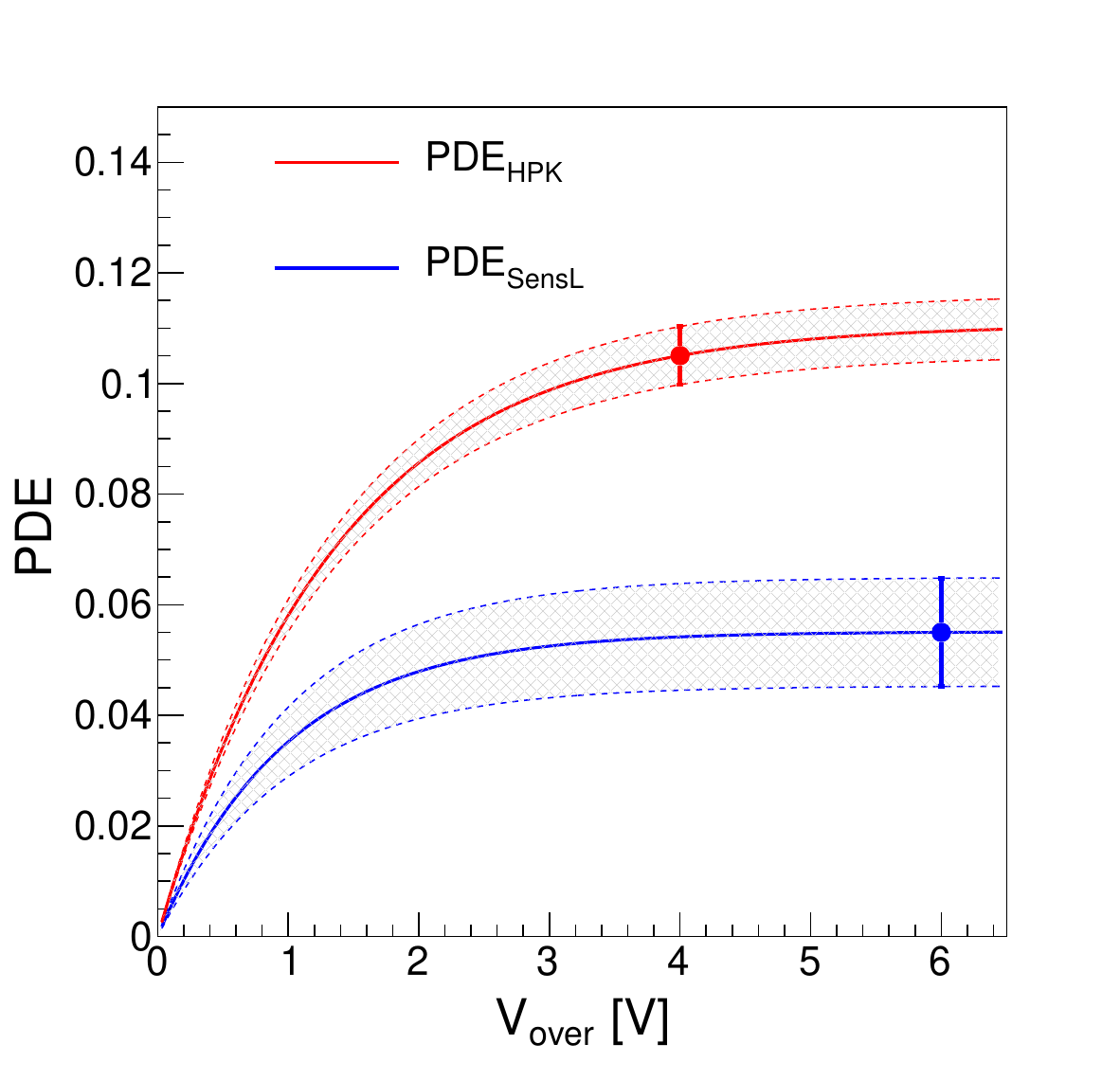}}
\caption{\label{PDE result} PDE versus $V_{over}$ for two SiPM samples for eCT photons. 
The functional form is defined by Eq.~\ref{PDE_420nm function}, with the required parameters listed in Table.~\ref{PDE parameter}.}
The point with error bar on the curves corresponds to the calculated PDE based on data from Fig.~\ref{Sensl_pde_pect}. And the shadows correspond to the uncertainty of PDE functions 
\label{PDE}
\end{figure}

PDE as a function of $V_{over}$ for both SiPM samples was obtained from the calculation and $\alpha$ from the fitting, as depicted in Fig.~\ref{PDE}.
The shadow areas in the figure represent the previously mentioned calculation results:  PDE=5.50$\pm$0.98\% for the SensL SiPM at $V_{over}$=6V and PDE=10.5$\pm$0.53\% for the HPK VUV4 SiPM at $V_{over}$=4V. 
Once the PDE function is confirmed, the value of $PDE_{max}$ can be obtained automatically. The results can also be found in Fig.~\ref{PDE}.
All parameters are included in Table.~\ref{PDE parameter}.

	\begin{table}[htb]
	    \caption{ \\ Parameters defined in Fig.~\ref{PDE_420nm function}. $V_{bd}$ for SiPM samples was measured under 76~K dark environment.} 
		\centering
		\normalsize
		\begin{tabular}{cccc}
		\hline
		SiPM type & $V_{bd}$ &  $PDE_{max}$ & $\alpha$\\ \hline
		HPK SiPM & 42.25 &  11.1$\pm$0.56~\% & 3.18$\pm$ 0.37$\times$10$^{-2}$\\
		SensL SiPM & 20.58 & 5.51$\pm$0.98~\% & 4.77$\pm$ 1.17 $\times$ 10$^{-2}$ \\
		\hline 
		\end{tabular}
	    \label{PDE parameter} 	
	\end{table}

\subsection{$P_{eCT}$ result}

\begin{figure}[t]
\centerline{\includegraphics[width=3.5in]{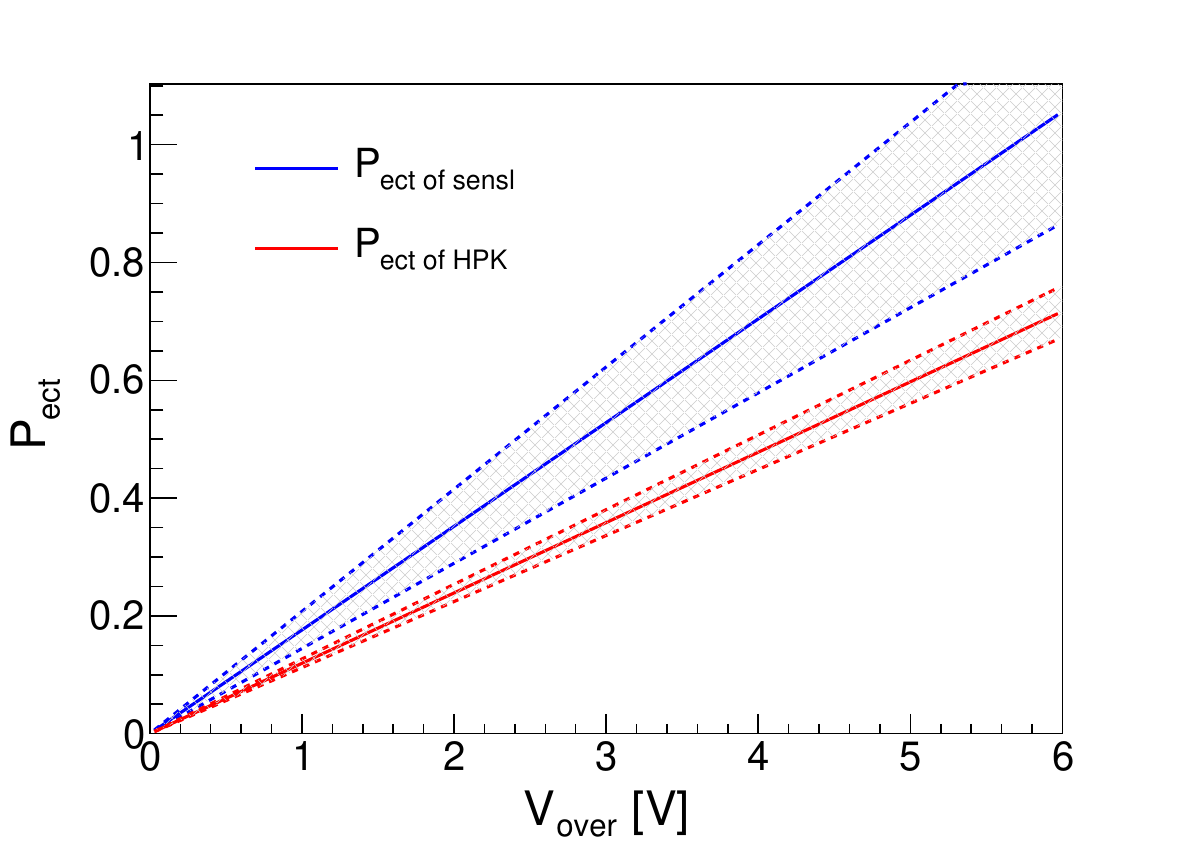}}
\caption{\label {p_ect}   $P_{eCT}$ functions with $V_{over}$ for two kinds of SiPM separately. The values of parameter $k$ are derived from the fitting results in Fig.~\ref{Fitting Result}.
The parameter $k$ for Sensl J-60035 SiPM was determined to be 1.76$\pm$0.31$\times$10$^{-1}$V$^{-1}$, with a 70.4\% probability of emitting at least one photon at the typical operating voltage of 4V. Similarly, the parameter $k$ for HPK VUV4 13370-6050CN SiPM was found to be 1.12$\pm$0.07$\times$10$^{-1}$V$^{-1}$, with a 44.8\% probability of emitting at least one photon at the typical operating voltage of 4V.
}
\end{figure}

Finally, parameter $k$, which corresponds to the number of eCT photon per avalanche, can be estimated since $k\times PDE_{max}$ was obtained in the fitting and  $PDE_{max}$ was achieved in Sec.~\ref{separating k and PDE}. 
The value of $k$ for the HPK VUV4 SiPM is shown in Fig~.\ref{p_ect}, which depicts the $P_{eCT}$ function versus $V_{over}$ using the estimated parameter $k$. 
The uncertainty of $k$ arises from the error in the parameter $k \times PDE_{max}$ from the fitting and the error propagated from $PDE_{max}$ obtained in Sec.~\ref{separating k and PDE}: 5\% for HPK VUV4 SiPMs and 17.8\% for SensL SiPMs. 
The shaded areas in Fig~.\ref{p_ect} represent the uncertainty of $P_{eCT}$ estimation. 
With the knowledge of the gain of SiPM samples, which are 5.80$\times$10$^6$ for HPK VUV4 13370-6050CN SiPM ~\cite{Hamamatsu2} and 1.93$\times$10$^6$ for Sensl J-60035 SiPM~\cite{PDiCT} at $V_{over}$=4V, the average number of photons per electron can be easily calculated. 
The results are included in Table.~\ref{PectResult}.

	\begin{table}[htb]
	    \caption{ \\  Number of eCT photons per electron.}
		\centering
		\normalsize
		\begin{tabular}{cc}
		\hline
		SiPM type & eCT photon number per e$^-$\\ \hline
		HPK SiPM & 7.66$\pm$0.48$\times$10$^{-8}$   \\
		SensL SiPM & 3.65$\pm$0.64$\times$10$^{-7}$   \\
		\hline 
		\end{tabular}
	    \label{PectResult} 	
	\end{table}

\section{Discussion}

In Sec.~\ref{sec:section4}, the probability of emitting a single photon per avalanche has been estimated for two types of SiPMs in our experimental setup. 
The eCT ratio calculated in the paper represents a statistical average and does not quantitatively demonstrate the angular acceptance.
In principle, the eCT probability depends on the specific detector configuration, resulting that the refractive index of the coupling between the SiPM and the detector could significantly impact the probability. 
A similar refractive index could reduce the likelihood of secondary photons reflecting back to the pixel that generated them. 
Consequently, iCT photons would be converted into eCT photons as a result of the change in the photon propagation medium. 
For example, a specific face-to-face setup would suggest a higher probability of photons emitted by a SiPM reflecting back onto the other SiPM, potentially resulting in a higher iCT ratio and a lower eCT ratio.
Utilizing probability as an approximation method is common practice. However, the exact number of photons emitted in each avalanche remains unknown. Moving forward, our focus will be on refining the measurement techniques and striving to incorporate precise numerical values for a more accurate depiction, including the calculation of the average number of photons emitted per avalanche.

An interesting byproduct of our method was the estimation of the PDE.
Currently, there are no studies on the SiPM PDE for the eCT spectrum. A flaw of this method is the normalized eCT spectrum necessary for estimating the PDE of the SiPM.  But, in practice,relative eCT spectrum is easy to be obtained in the experiment, and can be compared with the measurement of absolute eCT photon distribution as a function of wavelength.

\section{Conclusion}

This paper introduces a novel method to measure the probability of emitting one eCT photon per avalanche as a function of the SiPM overvoltage.
Two types of SiPM were explored.
The relationship between $P_{ect}$ and $V_{over}$ is a linear function with a single coefficient, denoted as $k$.
We found $k$ to be 1.76$\pm$0.31$\times$10$^{-1}$V$^{-1}$ for Sensl J-60035 SiPM and 1.12$\pm$0.07$\times$10$^{-1}$V$^{-1}$ for HPK VUV4 13370-6050CN SiPM.


\end{document}